\documentstyle[amsfonts,epsfig,12pt]{article}

\def\mypagenumber{1}

\def\myend{\end{document}}

\normalsize

\newcounter{sxn}

\newcounter{axn}

\date{}

\newdimen\mybaselineskip
\newcommand{\beeq}{\begin{equation}}
\newcommand{\eneq}{\end{equation}}
\newcommand{\be}{\begin{eqnarray}}
\newcommand{\ee}{\end{eqnarray}}
\newcommand{\bpic}{\begin{picture}}
\newcommand{\epic}{\end{picture}}

\def\la{\raise.16ex\hbox{$\langle$} \, }
\def\ra{\, \raise.16ex\hbox{$\rangle$} }

\def\psibar{ \psi \kern-.65em\raise.6em\hbox{$-$} }
\def\mbar{ m \kern-.78em\raise.4em\hbox{$-$}\lower.4em\hbox{} }

\def\n@space{\nulldelimiterspace=0pt \mathsurround=0pt }
\def\huge#1{{\hbox{$\left#1\vbox to 20.5pt{}\right.\n@space$}}}

\def\myskip{\noalign{\kern 8pt}}
\def\myeqspace{\noalign{\kern 10pt}}

\def\boxit#1{$\vcenter{\hrule\hbox{\vrule\kern3pt
    \vbox{\kern3pt\hbox{#1}\kern3pt}\kern3pt\vrule}\hrule}$}
\def\bigbox#1{$\vcenter{\hrule\hbox{\vrule\kern5pt
     \vbox{\kern5pt\hbox{#1}\kern5pt}\kern5pt\vrule}\hrule}$}

\def\ignore#1{{}}

\begin{document}

\bibliographystyle{unsrt}
\footskip 1.0cm

\thispagestyle{empty}
\setcounter{page}{\mypagenumber}

             
\begin{flushright}{
OUTP-01-39-P\\}

\end{flushright}

\vspace{2.5cm}
\begin{center}
{\LARGE \bf {Fluctuations of a (Tensionless) Brane World}}\\ 
\vskip 1 cm
{\large{Bayram Tekin}}$^{a,}$\footnote{e-mail:~
tekin@thphys.ox.ac.uk}\\
\vspace{.5cm}
$^a${\it Theoretical Physics, University of Oxford, 1 Keble Road, Oxford,
OX1 3NP, UK}\\

\end{center}

\vspace*{2.5cm}


\begin{abstract}
\baselineskip=18pt
We compute the quantum fluctuations of a 3-brane with tension, energy density and stiffness.
As a result of the fluctuations there are induced forces between massive objects living on the brane.
We study various limiting cases of the induced potential between 2 and 3 massive objects.
One quite interesting finding is that for tensionless brane world there are universal 
(mass independent) $1/r^3$ forces between the objects on the brane. These forces are
in principle measurable. 
\end{abstract}
\vfill

 
\newpage



\normalsize
\baselineskip=22pt plus 1pt minus 1pt
\parindent=25pt

The idea \cite{rubakov} that our universe is a brane-like object in a higher dimensional 
space with `large' or `small' extra spatial dimensions has not yet 
been shown to contradict the standard model and gravity. 
It is therefore of some importance to make testable predictions using brane-world pictures.
Unfortunately we are stuck on the brane and we lack the knowledge of how the full theory 
(brane + bulk) looks like. One can certainly build various models using M-theory \cite{witten,lukas}
but in principle one would like to test the brane-world idea without knowing the
the proper microscopic theory. Thus it would certainly be interesting to find  
some universal (model independent) predictions of the idea of a brane-world.

One interesting possible observable effect of living on a hyper-surface embedded in a
higher dimensional space is the modification 
of gravity at both small and large distances  \cite{arkani,antoniadis,randall,ross,
gregory,dvali}. Even though this is certainly an interesting observable effect 
of brane world idea; it was pointed out in \cite{tekin} that any observed modification of 
gravity at distances smaller than $ 0.1 mm $ would not necessarily show the
existence of extra dimensions since one could explain modifications of gravity at sub-millimeter
scales by staying in four dimensions and adding a higher
dimensional operator  $( \alpha R^2)$ to Einstein-Hilbert action.
Low energy limit of $  G^{-1} R +\alpha R^2$ theory in four dimensions
coincides with the low energy limit of higher dimensional theories.
Current experimental tests of gravity are compatible with a non-zero $\alpha$.

An other possibility in probing the extra dimensions is to study the observable
effects of the fluctuations of the brane in the bigger space. This question was
addressed in various works \cite{bando, kugo, creminelli, dobado,donini}.
In particular it was shown in \cite{bando}  that brane fluctuations remedy the ugly tree-level 
divergences that arise as a result of the Kaluza-Klein modes -matter interactions in the theories
where gauge fields are allowed to live in the bulk with some compact dimensions.
Brane fluctuations render the two charged particle scattering finite by 
suppressing the effect of higher Kaluza-Klein modes. One important result that comes out from the
above works is that the {\it{tension}} of the brane world {\it{cannot}} be arbitrarily small
since the fluctuations of the brane have some observable consequences which are proportional to
the negative powers of the tension. Specifically  as a result of the fluctuations there is 
an induced potential energy between two massive particles which is of the form \cite{kugo}
\be
V(r) =  - {3 (D-3)\over 2^7 \pi^3} {m_1 m_2 \over \tau^2 r^7} 
\label{pot1}
\ee
where $\tau$ is the tension of the brane-world. Clearly $\tau$ cannot be made too small, otherwise
Eqn. (\ref{pot1}) would dominate over gravity effects. In the literature one usually finds
that the tension of the brane world is taken to be around  $1 (TeV)^4$. 
Comparing the tension to the energy density $\epsilon$ of the brane we have
\be
 {\tau \over \epsilon} \sim { \mbox{TeV}^4 \over 10^{-12} \mbox{eV}^4} = 10^{60} 
\ee
Considering that for Nambu-Gotto branes one expects $\tau = \epsilon$; 
the large number $10^{60}$ is the usual `cosmological constant' problem 
which needs to be explained.

In this paper I will re-consider the transverse fluctuations of the brane world.
But my approach differs from the above mentioned works in the way that I use a different 
effective action for the brane.
I will assume that the brane has an  energy density ($\epsilon$) 
and stiffness ($\mu$)  in addition to its tension $\tau$.
Without the knowledge of the full theory it is conceivable that the {\it{effective action}} 
of the brane that describes its motion in the extra dimensions is not in the Nambu-Gotto
form but in a more general form. In fact there are examples of non Nambu-Gotto type behavior 
in similar objects: cosmic gauge strings. If there are small scales to the string 
(lets say some wiggliness ) then the  tension of a cosmic string is decreased whereas the 
energy density is increased yielding $\tau \neq \epsilon $ \cite{vilenkin,carter,kim}.
 I willstudy the observable
non-relativistic effects (low energy theory) therefore it is sufficient to consider a non-relativistic
brane action. But as far as phenomenology is concerned we do not loose much when
we work with the non-relativistic version. One very important consequence of our approach is that
we can consider the fluctuations of a brane with `small' tension and even we will consider 
{\it{tensionless}} 
brane world without having a divergent `fifth-force' which is given by Eqn. (\ref{pot1}). On the other 
hand in the limit of Nambu-Gotto brane we will reproduce the formula Eqn. (\ref{pot1}).

Our main idea is that we do not know what the full theory is and how the brane action is induced.
All we know is that we have standard model fields on a flat 3-brane which fluctuates in (D-3)
transverse dimensions. In a different context this problem was studied by 
D`Hoker et al \cite{kanev} some time ago and I will make use of their computations.  
The action of a  d-brane which describe the fluctuations in $(D-d)$ directions is
\be
S = \int dt \int d^d x {1\over 2} \bigg \{ \epsilon \biggl({\partial
\varphi^m\over \partial t}\biggr)^2 - \tau (\vec\nabla \varphi^m)^2 - \mu
(\vec\nabla\cdot\vec\nabla\varphi^m)^2 \bigg \}
\ee
where, once again, $\epsilon$ is the energy density, $\tau$ is the tension and $\mu$ is the stiffness
of the brane. $\varphi^m(t,\vec{x})$ is the position of the brane.
We will put freely moving uncharged objects on the brane. For $N$ objects (we do not specify what
these objects are yet. They can be as small as particles or as large as galaxies )  
the coupling to the brane is through the kinetic term,
\be
S_m =  \int dt \int d^d x \sum_{j=1}^N {1\over 2} m_j \biggr ( {\partial\varphi^m\over\partial t}\biggr)^2 \delta^d
(\vec x - \vec x_j)
\ee
By considering a flat brane we assume that we switch of the gravitational attraction between the
masses. In fact we should keep in our mind that gravity is the dominating force among neutral objects
and what ever interaction we will find should be a correction to gravity.
Following \cite{kanev} the vacuum energy of the theory is defined as
\be
-E = (D-d) \lim_{T\to\infty} {1\over 2T}~e^{\displaystyle{-{\sum_{j=1}^N}
{m_j\over 2} \int_{-T}^{T} dt \biggl( {\partial\over\partial t}
{\delta\over
\delta J(\vec x_j ,t)}\biggr)^2}} Z[J]\bigg\vert_{J=0\atop{\rm connected}}
\ee
where the generating functional and the propagator are
\be
Z[J] = e^{{1\over 2} \int d^{d+1} x\int d^{d+1} y J(x) \Delta_F (x-y)
J(y)}
\ee
\be
\Delta_F (x) = \int {d^{d+1} q\over (2\pi)^{d+1}}~~ {e^{iq\cdot x}\over
\epsilon
q_0^2 + \tau \vec q\cdot \vec q + \mu (\vec q \cdot \vec q)^2}\ .
\ee
Considering the proper diagrams one obtains the Coleman-Weinberg potential {\cite{kanev}}
\be
E = (D-d) \int_0^\infty {dw\over 2\pi} \mbox{log}\,\mbox{det}~(1+w^2 M)
\ee
where $M$ is the $N \times N$ matrix of elements
$M_{jk} = m_j G(w,\vert\vec x_j -
\vec x_k \vert)$, and:
\be
G(w,\vert \vec x \vert) =
 \int {d^d q\over
(2\pi)^d} {e^{i\vec q\cdot\vec x}\over \epsilon w^2 + \tau \vec q\cdot\vec q +
\mu (\vec q\cdot\vec q)^2}\ .
\label{green}
\ee
For the case of a 3-brane up to a divergent constant
the potential energy between two massive particles is

\be
V (r)= (D-3) \int_0^\infty {dw\over 2\pi}  \mbox{log}\left[ 1- {m_1m_2w^4
G(w,r)^2\over (1+m_1w^2G(w,0))(1+m_2w^2G(w,0))} \right]
\label{pot2}
\ee
Carrying out the integral  Eqn. (\ref{green}) for a 3-brane we get 
\be
G(w,r) = {1\over 4\pi r \sqrt{\tau^2 - 4\mu\epsilon \omega^2 }} \left \{ e^{{r\over \sqrt{2\mu}}\sqrt{ \tau -
\sqrt{\tau^2 - 4\mu\epsilon \omega^2}}} -   e^{{ r\over \sqrt{2\mu}}\sqrt{ \tau +
\sqrt{\tau^2 - 4\mu\epsilon \omega^2}}} \right \}
\ee

Leaving aside the full theory with non-zero stiffness and non-zero tension; let us compute what would 
be the induced potential in the two extreme cases of zero stiffness or zero tension.

{\underline{{\bf{Brane with zero stiffness}}}}: 

Setting $\mu = 0$ the Green's function reads 
\be
G(w,r)= {1\over 4\pi\tau r} e^{- wr \sqrt{{\epsilon\over\tau}}}
\label{zerostiff}
\ee
It is clear that the potential  Eqn. (\ref{pot2}) between the masses vanishes in this case
since $G(w,0)= \infty$. This seems rather boring. We can do a little better by introducing
a UV cut off, $l_P$ and not allow  $r = 0$.~\footnote{ As noted in \cite{kanev} a small stiffness
can provide such a cut off and one can take $l_p^2 = {\mu \over \tau}$. But here we merely introduce 
some small cut off, which we will take  to be the Planck length.} Doing so one obtains the following
potential at large distances (or small masses)
\be
V(r) = - {3 (D-3)\over 2^7 \pi^3} {m_1 m_2 \over \epsilon^2 r^7} ({\tau \over \epsilon})^{1\over 2}  
\hskip 2 cm \mbox{If}\hskip 1 cm     r^2 \gg  {m_{1,2}\over 4\pi \epsilon l_P }  
\label{pot3}
\ee
and small distances ( or large masses) 
\be
V(r) = - {(D-3)l_P^2 \over 4 \pi r^3} ({\tau \over \epsilon})^{1\over 2}  
\hskip 2 cm  \mbox{If}\hskip 1 cm   r^2 \ll  {m_{1,2}\over 4\pi \epsilon l_P }  
\label{pot3b}
\ee
First thing to notice is that in the in the 
Nambu-Gotto brane world limit ($\tau =\epsilon$) the large distance limit of the potential Eqn. (\ref{pot3}) 
coincides with Eqn. (\ref{pot1}) as promised.  
On the other hand in small distance limit, Eqn (\ref{pot3b})  defines a universal
attractive potential independent of the masses. 

If we take $l_P = 1.6\times 10^{-33}$cm and $\epsilon = 10^{-12} (\mbox{eV})^4$ then we can see that we are necessarily in the small distance limit. Therefore fluctuations of a brane with zero 
stiffness induce a mass independent
attractive potential between all the objects in the brane (as long as they are not strictly massless).
But since this potential is proportional to $l_P^2$ it is not observable.

{\underline{{\bf{Tensionless brane world}}}} : $\tau = 0 $

In this case one has
\be
G(w,r)={e^{-r \sqrt{{w\over 2}\sqrt{{\epsilon\over \mu}}} }\over 4\pi r w
\sqrt{\epsilon\mu}}~\sin \biggl( r \sqrt{{w\over 2}\sqrt{{\epsilon\over\mu}}}
\biggr )
\ee
and the induced potential energies are  computed to be
\be
V(r) = - {15 (D-3)\over 2^5\pi^3} {m_1 m_2 \over \epsilon^2 r^8} ({\mu \over \epsilon})^{1\over 2}  
\hskip 2 cm  \mbox{If}\hskip 1 cm    r^2 \gg  {m_{1,2}\over 4\pi \epsilon l_P }  
\label{pot3c}
\label{zerotension}
\ee
and 
\be
V(r) = - {0.86 (D-3)\over 2\pi} {1 \over r^2} ({\mu \over \epsilon})^{1\over 2}  
\hskip 2 cm  \mbox{If}\hskip 1 cm  r^2 \ll  {m_{1,2}\over 4\pi \epsilon l_P }  
\label{pot4}
\ee
Therefore for tensionless brane-world  once again we obtain a  
universal (in the lowest order)  potential between the masses in the small $r$ limit.
Observe that small 
$r$ really means smaller than the radius of the universe, since  
$r^2 \gg  {m_{1,2}\over 4\pi \epsilon l_P }$ is hard to satisfy for reasonable masses. 
 
Eqn. (\ref{pot4}) predicts a measurable modification of Newton's law. 
Non-observation of this universal potential, say  in the solar system,
puts a limit on the stiffness of the brane world. At larger scales one can still
hope to see the effect of  Eqn. (\ref{pot4}).

{\underline{ {\bf{Interaction of 3 masses}}}}

Let us now compute the induced potential energy between tree massive objects that are 
embedded in the brane.
For simplicity I will assume that all these masses are at the vertices of an equilateral triangle.
Up to a divergent constant the potential energy is

\be
&&V (r) =(D-3) \int_0^\infty {dw\over 2\pi} \nonumber  \\ 
&&\mbox{log}\left[ 1 + 
{m_1 m_2 m_3 w^6 G(w,r)^2 (2 G(w,r) - 3 G(w,0) ) - 
(m_1 m_2 + m_1 m_3 + m_2 m_3)w^4 G(w,r)^2 \over (1+m_1 w^2G(w,0))(1+m_2 w^2 G(w,0))(1+m_3 w^2G(w,0)) } \right] \nonumber
\ee
In the zero stiffness ($\mu = 0$ ) limit one has 
\be
V(r) &&= - {3(D-3)\over 2^7\pi^3} {(m_1 m_2 + m_1 m_3 + m_2 m_3) \over \epsilon^2 r^7} ({\tau \over \epsilon})^{1\over 2} \nonumber \\
&& - {135(D-3)\over 2^{10}\pi^4} {m_1 m_2 m_3 \over \epsilon^3 r^9 l_P} ({\tau \over \epsilon})^{1\over 2}  
\hskip 0.5 cm \mbox{If}\hskip 0.5 cm  r^2 \gg  {m_{1,2,3}\over 4\pi \epsilon l_P } 
\ee
Therefore for large distances there is an attractive three body potential 
induced on the brane in addition to the dominant two body potential. 
For large masses the story is a little different. The potential is again
universal as in the case of two masses but now it is coefficient is three times larger.
\be
V(r) = - {3 (D-3) l_P^2 \over 4 \pi r^3} ({\tau \over \epsilon})^{1\over 2}  
\hskip 0.5 cm  \mbox{If}\hskip 0.5 cm  r^2 \ll  {m_{1,2,3}\over 4\pi \epsilon l_P }
\ee
Once again zero stiffness limit does not give a measurable prediction 
since the relevant potential energy is tiny.

In the tensionless brane case we have

\be
V(r) &=& - {15 (D-3)\over 2^5\pi^3} {m_1 m_2 +m_1 m_3 +m_2m_3 \over \epsilon^2 r^8} ({\mu \over \epsilon})^{1\over 2} \nonumber \\
&&   {28.6(D-3)\over \pi^4} {m_1 m_2 m_3 \over \epsilon^3 r^{11}} ({\mu \over \epsilon})^{1\over 2} 
\hskip  0.5 cm  \mbox{If}\hskip 0.5 cm  r^2 \gg  {m_{1,2,3}\over 4\pi \epsilon l_P }
\ee
and at short distance the relevant integral is
\be
V(r) = {D-3 \over 2\pi} \int_0^\infty d w \, \mbox{log}\left [ 1 - 3 {G(w,r)^2\over G(w,0)^2} + 
2{G(w,r)^3\over G(w,0)^3} \right ]    
\ee
which gives~\footnote{ The coefficient is determined by the integral :  
$\int_0^\infty d x \, \mbox{log}\left [ 1 - 3 {e^{-2x} \sin^2 x\over x^2}  
+2{e^{-3x} \sin^3 x\over x^3} \right ] = - 0.52 $ }     
\be
V= -{D-3\over \pi r^2}  ({\mu \over \epsilon})^{1\over 2}
\hskip 1 cm  r^2 \ll  {m_{1,2,3}\over 4\pi \epsilon l_P }
\label{pot7}
\ee
The relevant limit is the Eqn. (\ref{pot7}) for which $r$ is smaller than the horizon.

To conclude I have shown that considering the quantum fluctuations of generic brane action
with tension, energy density and stiffness one can study the observable effects of 
extra dimensions. One important consequence of this computation is that we are allowed to
consider tensionless branes as opposed to the usual practice of taking the tension
very high. In a sense we have solved the large tension problem by going to 
non-relativistic theory.
We have also shown that Nambu-Gotto limit $\tau = \epsilon = 10^{-12} (\mbox{eV})^4$ is possible
without introducing a large force. 
In the tensionless brane world picture, if there is a non-zero stiffness to the brane, fluctuations 
of the brane induce mass independent potentials which are in principle measurable.
I have also considered the induced potential between 3 massive objects.
Once again there are universal potentials in the tensionless brane limit.
In principle one can compute the finite temperature effects by using the methods of \cite{kanev}
but I shall not dwell on this here.

I would like to thank A. Ibarra and S. Sarkar for useful discussions. 
This work is supported by  PPARC Grant PPA/G/O/1998/00567.

\myend